# Graph representation of context-free grammars

Alex Shkotin
Department of software engineering,
Computing Centre RAS,
Moscow, Russia
ashkotin@ccas.ru

## Abstract

In modern mathematics, graphs figure as one of the better-investigated class of mathematical objects. Various properties of graphs, as well as graph-processing algorithms, can be useful if graphs of a certain kind are used as denotations for CF-grammars. Furthermore, graph are well adapted to various extensions (one kind of such extensions being attributes).

This paper describes a class of graphs corresponding to CF-grammars – the decision-making graphs (henceforth abbreviated as DMG).

## Construction DMG from a CF-grammar

Some designations:
- An uppercase Latin letter, optionally followed by digits, designates a nonterminal (for example: A, B, X, S1).
- A lowercase Latin letter designates a terminal (for example: a, b, c).
- A lowercase Greek letter designates a chain of nonterminals and/or terminals. As a special case, ε always designates an empty chain.

Supposing a CF-grammar (CFG) is given (see for example [1]), it is assumed that the CFG in question has the following properties.

**There is no identity** – the grammar does not contain rules of the form A→A (the uselessness of such rules is obvious).

**Lexical nonterminals** – nonterminals may exist in the grammar that not do appear in the left part of any rule. This situation is typical for grammars describing only the syntax of the language; such nonterminals are usually starting symbols of regular grammars describing the lexical structure of the language.

### Reduction of rules to a form convenient for graph construction

Let's break the set of derivation rules specified by the grammar into two disjoint subsets according to the properties of the symbol appearing in the left part of the each rule:

**The zone of non-determinism** (ZnD) – this subset will contain all rules such that the symbol in their left part appears in more than one rule.

**The zone of a determinism** (ZD) – this subset will contain all rules such that the symbol in their left part appears in one rule only.

An additional requirement for the grammar in question is:

**There is no bad infinity** – ZD shall not contain any rules of the form A→αAβ. Rules in ZD specify a unique way of substitution for A, and rules of the form A→αAβ mean that the derivation of any chain containing A will never terminate. In real grammars, the author of language initially sees to it that rules of the form A→αAβ never appear in ZD.

### Reduction of rules in ZnD to a "one-to-one" form

**Addition of missing nonterminals**: Let B→α be some rule from ZnD such that the length of α is not equal to 1. To perform a reduction of this rule, we shall introduce a new nonterminal (let's call it X) not appearing anywhere in the grammar. The rule shall then be transformed to a pair of rules (B→X; X→α). Thus B→X remains in ZnD, and X→α goes to ZD.

After the above transformation is applied to all ZnD rules, we come to a situation where for each nonterminal B that appears in the left part of more than one rule all derivations contain exactly one symbol in their right part, whether terminal or not.

**Definition**: A grammar whose ZnD has been reduced to "one-to-one" form shall be referred to as a determined not determined (DnD) grammar.

### Assigning types to grammar symbols

Now let's assign a letter from the alphabet {0&!} – "type of a symbol" – to each symbol used in the grammar using the following rules:
- Terminals are assigned type "0".
- Nonterminal that appear in the left part of more than one rule are assigned type "!" (these shall be referred to as **OR-nonterminals**).

- Nonterminal that appear in the left part of only one rule are assigned type "&" (these shall be referred to as **AND-nonterminals**).
- Nonterminal that do not appear in the left part of any rule are assigned type "0" (lexical nonterminals).

Note: It is assumed that the author of the language has intentionally left some nonterminal without derivation rules, these being lexical nonterminals, and grammar describing only the syntax of the language.

*DMG construction algorithm*

DMG is a directed ordered graph without loops with typed nodes. Each node of the graph is labeled with the symbol of the grammar that node represents.

To construct a DMG from a DnD grammar:
1. Create a node for each symbol of the grammar (whether terminal or nonterminal) and label the node with that symbol. Labeling a node with a symbol implicitly labels that node with the type of that symbol (so we will refer to a node labeled with an AND-nonterminal as an **AND-node**, and a node labeled with an OR-nonterminal as an **OR-node**).
2. For each OR-node labeled with some symbol X, create edges from that node to all nodes corresponding to symbols that appear in the right parts of derivation rules where X appears as the left part. The edges are ordered arbitrarily.
3. For each AND-node labeled with some symbol X, create edges from that node to nodes corresponding to all symbols that appear in the right part of the only derivation rule for X. The number of such edges is equal to the length of the rule's right part. Each edge represents an inclusion of some symbol into the derivation of X and is labeled with the ordinal of that inclusion. If the right part of the derivation rule for X is empty, there will be no outgoing edges from the corresponding DMG node.

The graph is constructed.

*Example: A grammar and its graph*

Let's consider Example 4 of [1]:
S → aSc ! B
B → bBc ! ε

Reducing the grammar to a DnD results in:
S → S1 ! B
B → B1 ! B2
S1 → aSc
B1 → bBc
B2 → ε

The corresponding DMG looks as follows on Fig. 1:

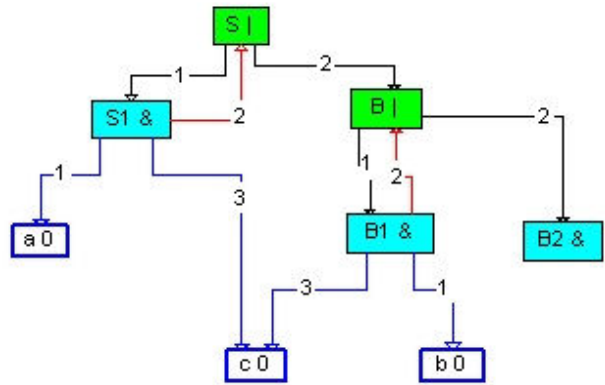

Figure 1: DMG example

(The colors on the DMG are used to facilitate viewing big graph and have no other significance.)

**Small pleasures**
- Each symbol of the grammar is represented by one and only one node.
- Using incoming edges, it is easy to see in which rules a given symbol appears in the right part. This is especially interesting for the terminals, these being the keywords of the language.
- The DMG contains two cycles – an attribute of a "true" language.

### DMG in itself - basic properties of DMG

DMG is a directed graph without loops where edges starting at each node are ordered and nodes attributed in such a way that:
- Nodes are uniquely marked with symbols of grammar;
- Nodes are assigned types {&0!} so that:
  - !-node has at least 2 outgoing edges.
  - There are no parallel outgoing edges from an !-node.
  - There are no outgoing edges from 0-nodes.

It is clear that to check for the correctness of types in a DMG it is necessary to check properties of edges outgoing from !- and 0- nodes.

### The accepted decisions tree

For a DMG there exists an analogue of a derivation tree – a tree of accepted decisions (henceforth abbreviated as TAD).

Given a DMG, we shall first shall define an "almost TAD" (aTAD) as follows:
- An isolated node marked S is an aTAD. (S is a starting nonterminal of a CFG.)

- If a node N1 is a leaf of an aTAD, its type is "!" and its label is L1, then for <u>some</u> edge A1 outgoing from the DMG node with the label L1 do the following: 1) create a new edge A2 with the same number outgoing from N1; 2) create a new node N2; 3) attach the incoming end of A2 to N2 and 4) mark N2 with the same symbol as that marking the incoming end of A1. The result of such transformation is an aTAD.
- If a node N1 is a leaf of an aTAD, its type is "&" and its label L1, then for <u>each</u> edge A1 outgoing from the DMG node with the label L1 do the following: 1) create a new edge A2 with the same number outgoing from N1; 2) create a new node N2; 3) attach the incoming end of A2 to N2 and 4) mark N2 with its same symbol as that marking the incoming end of A1. The result of such transformation is an aTAD.
- There are no other aTADs except those that can be built using the rules 1..3 above.

Definition of TAD:

Given an aTAD in which all leaves have type "0" or type "&" (the latter - without outgoing edges in the DMG), replacing labels of "0"-nodes corresponding to lexical nonterminals with any allowable values for these nonterminal will result in a TAD.

An idea of TAD is simple – it records all decisions made in derivations of OR-nodes and all constructions in AND-nodes down to a terminal chain.

**The string of a formal language** is constructed as usual from the crone of a TAD and contains terminals and lexemes of the program. The only nuance is in that the crone of TAD can contain AND-nodes without outgoing edges (nodes forming an empty chain). Such nodes will be "gone" from the string.

Example:

Let's consider the grammar from the section "Derivations and syntax trees" of [1]:
(1) S → S + S
(2) S → 1
(3) S → a

Its DnD version is:
S → S1 ! 1 ! a
S1 → S + S

And on Fig. 2 is the DMG:

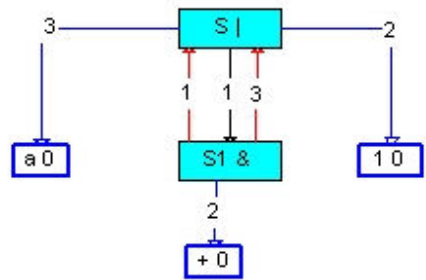

Figure 2: DMG example #2

The following TAD (on Fig.3) corresponds to the first tree on the specified page:

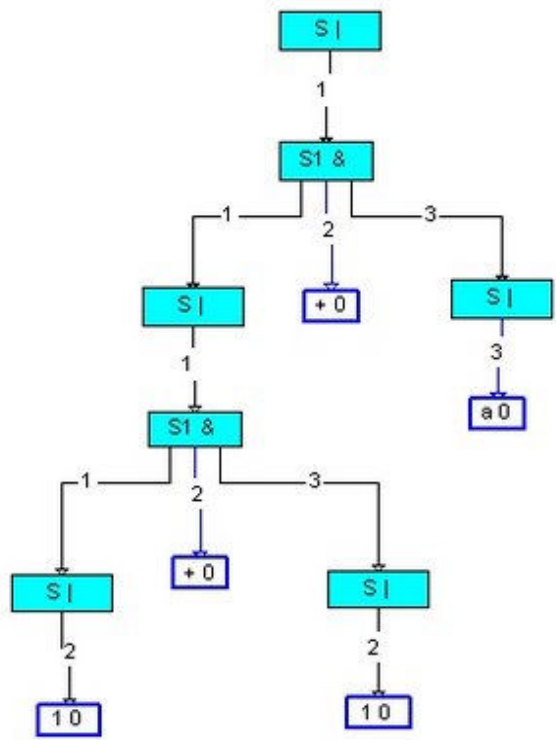

Figure 3: TAD example

The meaning of the TAD is simple enough: we have started with the node S and have decided to follow the edge "1", have reached the node S1 and have constructed its children. Now we have two more S-nodes. For the left one we have decided again to follow the edge "1", and for the right one to follow the edge "3", that results in the node "a", etc.

**Formal language on the graph**

A formal language can be defined without using TADs.

Let's define symbol chains derived from a node recursively as follows:
1. For a 0-node labeled with a terminal symbol, the derived symbol chain is that terminal symbol.
2. For a 0-node labeled with a nonterminal symbol, and given that α is a chain from the "predefined" (lexical) set assigned to that nonterminal, α can be derived from that node.
3. For an OR-node, and given that α can be derived from any of the node's children, then α can be derived from this OR-node.
4. For an AND-node, given that $α_i$ can be derived from a node located on the end of i-th outgoing edge, a chain concatenation of $α_1, α_2...α_N$ (where N is number of outgoing edges of the AND-node) can be derived from that AND-node (if an AND-node has no outgoing edges, then only an empty chain can be derived from it).
5. No other symbol chains can be derived except those specified by the rules 1..4 above.

The set of symbol chains that can be derived from a node is, naturally, equal to the set of symbol chains that can be derived from a symbol labeling the node.

Accordingly, the formal language defined by DMG is a set of terminal symbol chains that can be derived from a node marked with an initial symbol of the grammar (S).

The symbol chain derived from a given node will sometimes be referred to as a symbol chain of that node.

Remarks:
- Ordering edges of OR-nodes has no value.
- Labels of OR- and AND-nodes have no value. It provides a certain freedom in the choice of identifiers for nonterminal; the important part is in that DMG "contains" the language in the graph part, i.e. at the other level of abstraction. A graph illustrates the character of !&-nonterminals much better – they just name the corresponding node. It is also possible to look up node names in different natural languages and so on.

It is also interesting to determine why the author of a language has named some AND-nodes and did not name others – after all, we had to name all of them when reducing grammar to DnD.

### An abstract construction process

Using a symbol chain of a formal language as an example, it is possible to define the process of constructing in abstract terms. Specifically, to build a construction (a symbol chain of formal language) it is necessary to:
**Build the TAD**: construct TAD using DMG.
**Assemble the construction**: transfer parts of the symbol chain from the bottom of the TAD up, concatenating these parts in AND-nodes.

The result is a construction (symbol chain) assembled in the root S-node.

To find out what is done in AND-nodes when a construction is more complex than a chain of symbols, see d-schemes [2].

It is possible to design DMG directly, assuming that it sets some basis for creation of "constructions". Examples of "construction" are symbol chains of formal language and trees of the accepted decisions. Trees of the accepted decisions and their crones (symbol chains of a formal language) are directly set by the graph. In addition, we can also specify just what exactly we are building in AND-nodes. For example it is possible to build graphs, trees of a special kind, diagrams, etc. It is even possible to try and build programs, but that requires us to specify a construction of the program. It is unlikely to be a TAD and is certainly not a symbol chain of a formal language! As the first approximation it is possible to consider d-schemes.

### Discussion

Finding useless and inaccessible symbols (as well as many other things) turn out to be graph algorithms. It is also possible to search the graph for cycles as well as split the graph into two sub-graphs in such a way that all inter-sub-graph edges will only go in one direction (the latter corresponds to a "sublanguage"). Even the statistics trivial for the graph can be of much interest as far as the grammar is concerned. Among other things, the quantity of AND-nodes shows a number of <u>different</u> constructs possible in a program.

OR-nodes of a DMG can "behave" not just non-deterministically, but also interactively – requesting what alternative to choose. In this case, the user shall at least be made aware of what place of an aTAD he is currently in.

DMGs are equivalent CFGs and in this sense DMG can be ambiguous (different TAD can have an identical crone). Although parsing algorithms can work on a graph, the problem of building a TAD given its crone will have the same complexity. The only thing that is worth noting here is that if DMG is used to help the user in the construction of a program then we should keep not only the crone, but also the TAD, in order to avoid having to restore it later.